\documentstyle[twocolumn,seceq]{jpsj}
\input epsf.tex

\def\lsim{\mathop {\vtop {\ialign {##\crcr
$\hfil \displaystyle {<}\hfil $\crcr \noalign {\kern1pt \nointerlineskip }
$\,\sim$ \crcr \noalign {\kern1pt}}}}\limits}

\def\runtitle{Non-Collinear Magnetism due to Orbital Degeneracy and Multipolar Interactions}
\def\runauthor{Hiroaki {\sc Kusunose} and Yoshio {\sc Kuramoto}}

\title
{Non-Collinear Magnetism due to Orbital Degeneracy\\
and Multipolar Interactions}

\author
{
Hiroaki {\sc Kusunose}\footnote{E-mail: kusu@cmpt.phys.tohoku.ac.jp}
and
Yoshio {\sc Kuramoto}
}

\inst
{
Department of Physics, Tohoku University, Sendai 980-8578
}

\recdate
{
December 19, 2000
}

\abst
{
The origin of non-collinear magnetism under quadrupolar ordering is investigated with CeB$_6$ taken as a target system.
The mode-mixing effect among 15 multipoles is analyzed based on the Ginzburg-Landau free energy.
Then the lower magnetic transition temperature and the order parameters are derived within the mean-field approximation.
In the presence of pseudo-dipole-type interactions for the next-nearest neighbors, the observed pattern of non-collinear ordering is indeed stabilized for certain set of interaction parameters.
The stability of the phase III$^\prime$ in the magnetic field is also explained, which points to the importance of the next-nearest-neighbor octupole-octupole interaction.
Concerning the phase IV in Ce$_x$La$_{1-x}$B$_6$ with $x\sim 0.75$, a possibility of pure octupole ordering is discussed based on slight modifications of the strength of interactions.
}

\kword
{
orbital degeneracy, multipole moment, non-collinear magnetism, CeB$_6$, Ce$_x$La$_{1-x}$B$_6$, Ginzburg-Landau free energy, mode mixing
}

\begin{document}
\sloppy
\maketitle

%
%
\section{Introduction}
For the last decade, CeB$_6$ and Ce$_x$La$_{1-x}$B$_6$ have attracted great interest because of their exotic properties originating from orbital degeneracy.
Due to strong spin-orbit interaction of $f$ electrons, the entangled spin and orbital degrees of freedom show unusual responses against various external perturbations.

One of the mysteries was concerned with so-called phase II: (i) the inconsistency between NMR\cite{takigawa83} and neutron scattering\cite{effantin85,erkelens87} measurements on the pattern of the magnetic field induced antiferro (AF) magnetic moment and (ii) the increase of the transition temperature with increase of the magnetic field\cite{takigawa83,effantin85,fujita80}.
The latter has been interpreted theoretically in terms of the intersite interaction between the field-induced $\Gamma_2$-type octupole, i.e., $T_{xyz}$\cite{shiina97}, and the quadrupolar fluctuations suppressed by the magnetic field\cite{uimin96,kuramoto98}.
The former mystery has been resolved by reconsidering the NMR analysis, where the coupling to the induced octupole in the hyperfine interaction is taken into account\cite{sakai98}.
Through these studies, the AF-quadrupolar ordering has been identified as of $\Gamma_5$ type ($O_{yz}$, $O_{zx}$, $O_{xy}$).

In spite of these successful studies, understanding of phase III is still obscure.
A complicated magnetic structure with double ${\mib k}$, i.e., ${\mib k}_1=(1/4,1/4,1/2)$ and ${\mib k}_2=(1/4,-1/4,1/2)$ in units of $2\pi/a$ with $a$ being the lattice constant, has been established by neutron scattering\cite{effantin85,erkelens87}.
The non-collinear spin orientation has no gain of energy from the nearest neighbor exchange interactions.
Besides, the next-nearest neighbors give the same number of parallel and anti-parallel spins.
Thus, the double ${\mib k}$ structure is hardly stabilized by simple exchange interactions.

Furthermore, a new phase called IV was discovered recently in Ce$_x$La$_{1-x}$B$_6$ with $x\sim 0.75$, whose order parameter has not been identified yet.
The phase IV has prominent features: (i) a drastic softening of the transverse elastic constant $C_{44}$ ($\sim 30\%$)\cite{nakamura97}, (ii) almost no magneto resistance in contrast with other phases\cite{hiroi97}, and (iii) almost isotropic magnetic susceptibility in contrast with phases II and III\cite{tayama97}.
Although the magnetic susceptibility shows a cusp at the phase transition to the phase IV\cite{tayama97}, preliminary experiment of neutron scattering for powder samples has found no magnetic Bragg scattering\cite{kohgi}.
The systematic study of La substitution shows that the phase IV appears when the transition temperature $T_N$ to the phase III exceeds the AF quadrupolar transition temperature, $T_Q$.

Motivated by this situation, we consider it necessary first to investigate favorable conditions to stabilize the non-collinear orientation under the $\Gamma_5$-type AF quadrupolar ordering.
In this paper we analyze the mode-mixing effect among 15 multipole moments in the lowest $\Gamma_8$ quartet based on the Ginzburg-Landau (GL) free energy functional under the AF $O_{xy}$ ordering.
To account for the observed non-collinear ordering, we introduce the next-nearest neighbor interactions of a pseudo-dipole type\cite{sakai,kasuya66,yildrim95}, and explore possible mean-field solutions.
Based on the mean-field solutions, the stability of the phase III$^\prime$ for applied magnetic field along [111] is also explained.
This analysis points to the importance of the next-nearest-neighbor octupole-octupole interactions.
Then, we discuss a possibility of incommensurate $\Gamma_5$-type octupole ordering instead of the quadrupolar one with slight modifications of interactions.

This paper is organized as follows.
In \S2 we introduce the model Hamiltonian for the active multipoles of the $\Gamma_8$ quartet.
The GL free energy is derived microscopically from the given Hamiltonian through the path integral.
In \S3 we first discuss the limit of the large quadrupolar ordering temperature as compared with the N\'eel temperature.
Then, the possible mean-field solutions are presented for the nearest-neighbor and the next-nearest-neighbor interactions, respectively.
The summary is given in the final section.
Three appendices are given for details of calculation, in which we respect more general features of multipole systems than those in CeB$_6$.

%
%
\section{Model and Formalism}
\subsection{Active multipoles in $\Gamma_8$ subspace}
The four states in the $\Gamma_8$ level, which is the crystalline-electric-field ground state of Ce$^{3+}$ ion, are represented in terms of the basis $|J_z\rangle$ of $J=5/2$ as follows:
\begin{equation}
\left|1\pm\right)=\sqrt{\frac{5}{6}}\left|\pm\frac{5}{2}\right\rangle+\sqrt{\frac{1}{6}}\left|\mp\frac{3}{2}\right\rangle,\;\;\left|2\pm\right)=\left|\pm\frac{1}{2}\right\rangle.
\end{equation}
The entanglement of spin and orbital degrees of freedoms can be described most physically by multipole operators in accordance with the point-group symmetry.
To describe the multipole moments in a concise way, we introduce the pseudo-spin operators which are represented by two sets of Pauli matrices, ${\mib\sigma}$ and ${\mib\tau}$.
The latter operators act on the orbital partners, while the former on the Kramers pairs.
Namely, we have
\begin{equation}
\tau_z\left|m\pm\right)=(-1)^{m-1}\left|m\pm\right),\;\;\;\sigma_z\left|m\pm\right)=\pm\left|m\pm\right),
\end{equation}
with $m=1,2$.
The $x$ and $y$ components of pseudo-spins change one state to another in the $\Gamma_8$ subspace.
We can express the physical operators $X^A$ adapted to the point-group symmetry using the pseudo-spins\cite{shiina97}.
The capital letter $A$ specifies the set of indices in the irreducible representation $\Gamma$ with the multiplicity and the component $\gamma$.
We also assign a sequential number ($1\sim 15$) to $A$ for simplicity.
The multipole operators are summarized in Table I, where we have introduced linear combinations of $\tau^x$ and $\tau^z$ as
\begin{equation}
\eta^\pm=\frac{1}{2}\left(\pm\sqrt{3}\tau^x-\tau^z\right),\;\;
\zeta^\pm=-\frac{1}{2}\left(\tau^x\pm\sqrt{3}\tau^z\right).
\end{equation}
The subscript $u$ represents the odd property under the time reversal, and $g$ the even one.

Among these operators, $\tau^y$ in $\Gamma_{2u}$ has the same matrix element as the symmetrized product of $J^xJ^yJ^z$, and is regarded as a component of the octupole moment tensor.
On the other hand, the operators belonging to $\Gamma_{3g}$ and $\Gamma_{5g}$ describe the quadrupole operators.
The dipole operator ${\mib J}$ is decomposed into $\Gamma_{4u1}$ and $\Gamma_{4u2}$ under the cubic symmetry.
The magnetic moment ${\mib M}$ is given in units of the Bohr magneton by
\begin{equation}
{\mib M}={\mib X}^{4u1}+\frac{4}{7}{\mib X}^{4u2}.
\end{equation}
Note that dipole and a part of octupole operators have the same representation $\Gamma_{4u}$ under the point-group symmetry, and they are mixed with each other.
In this sense, we classify the third-rank tensors ${\mib X}^{4u2}$ into dipole moment operators.
The remaining representation $\Gamma_{5u}$ corresponds to pure octupole operators.

\subsection{Hamiltonian and pseudo-dipole interaction}
There is not much information about the form and range of the intersite interactions in CeB$_6$.
Therefore we proceed by assuming the simplest one and, if it turns out insufficient, add necessary ones in the next step.
Among various possible choices in the second step, we retain only the interaction which is essential in understanding the peculiar ordering pattern.
We assume that the intersite interactions between $i$ and $j$ are classified according to the representations of the cubic group.
The conduction electrons that give rise to the Kondo effect are not treated explicitly\cite{kusunose99a,kuramoto98a,kusunose99b}.
Then the Fourier transform of the nearest neighbor interaction for the 3D simple cubic lattice is given by
\begin{equation}
J_{\Gamma}({\mib q})_{\gamma\gamma'}=-J_{\Gamma}\delta_{\gamma\gamma'}j_0,
\label{nnint}
\end{equation}
with
\begin{equation}
j_0=\frac{1}{3}(\cos{q_x}+\cos{q_y}+\cos{q_z}).
\end{equation}
To realize the AF $\Gamma_{5g}$ order, $J_{5g}$ should be positive and the largest among all interactions.
The strength of the nearest neighbor interactions mediated by conduction electrons was discussed on the basis of the group-theoretical argument, where it was concluded that there are only two independent couplings under the AF quadrupolar ordering\cite{shiba99}.

It is evident that the above interaction alone does not stabilize the orthogonal arrangement of nearest-neighbor spins observed in the phase III.
Even though the above type of interaction is extended to next-nearest neighbors, the interaction does not stabilize the arrangement of spins at next-nearest neighbors; being either paralell or anti-parallel.
As we shall discuss in detail later, the following next-nearest neighbor interaction of the pseudo-dipole type is indispensable to account for the observed non-collinear spin orientation:
\begin{equation}
K^{\Gamma\gamma\gamma'}_{ij}=-K_{\Gamma}(\delta^{\gamma\gamma'}-3n^\gamma_{ij} n^{\gamma'}_{ij})/12,
\end{equation}
where ${\mib n}_{ij}$ is the unit vector across the next-nearest neighbor $i$ and $j$ sites.
The pseudo-dipole interaction is considered for three-dimensional odd representations, $\Gamma_{4u1}$, $\Gamma_{4u2}$ and $\Gamma_{5u}$.
Note that the positive (negative) $K_{\Gamma}$ favors (anti) parallel alignment along ${\mib n}_{ij}$.
In the {\mib q} space this corresponds to
\begin{equation}
K_{\Gamma}({\mib q})_{\gamma\gamma'}=-K_{\Gamma}
\left(\begin{array}{ccc}
j_x & j_{xy} & j_{zx} \\
j_{xy} & j_y & j_{yz} \\
j_{zx} & j_{yz} & j_z
\end{array}\right)_{\gamma\gamma'},
\label{pdint}
\end{equation}
where
\begin{eqnarray}
&&j_x=\frac{1}{6}\biggl[2\cos(q_y)\cos(q_z)-\cos(q_z)\cos(q_x)\nonumber\\
&&\;-\cos(q_x)\cos(q_y)\biggr],\;
j_{xy}=\frac{1}{2}\sin(q_x)\sin(q_y),
\end{eqnarray}
and other components are given by cyclic rotation.
The microscopic origin of the pseudo-dipole type interactions has been discussed by taking into account the RKKY or the superexchange interactions together with the $d$-$f$ exchange or the Hund's-rule coupling\cite{sakai,kasuya66,yildrim95}.

In actual systems, there should be other interactions than $J_{\Gamma}$ and $K_{\Gamma}$.
However we neglect the other interactions since they are irrelevant to giving rise to the observed spin structure in the phase III.
Consequently, the Hamiltonian used in this paper is given by
\begin{eqnarray}
H&&=-\frac{1}{2}\sum_{i\ne j}\sum_{AB}D^{AB}_{ij}X^A_iX^B_j \nonumber \\
&&=-\frac{1}{2}\sum_{\mibs q}\sum_{AB}D^{AB}_{\mibs q}X^A_{\mibs q}X^B_{-{\mibs q}},
\label{hamform}
\end{eqnarray}
with the interaction
\begin{equation}
{\sf D}_{\Gamma}({\mib q})=\left\{
\begin{array}{ll}
{\sf J}_{\Gamma}({\mib q}), & (\Gamma=2u,\;3g,\;5g),\\
{\sf J}_{\Gamma}({\mib q})+{\sf K}_{\Gamma}({\mib q}), & (\Gamma=4u1,\;4u2,\;5u).
\end{array}\right.
\end{equation}
The summation symbols for repeated indices will be omitted hereafter.

\subsection{GL free energy under AF $O_{xy}$ ordering}
The pseudo-spin representation is introduced just to reproduce the matrix elements of multipole operators.
Hence any approximation to decouple $\sigma$ and $\tau$ is not meaningful physically.
Instead we should consider on equal footing each fluctuation with a point-group symmetry.

To deal with coupled multipole fluctuations from the high-temperature side, we work with the path-integral representation of the partition function\cite{kuramoto98,negele}:
\begin{equation}
Z=\int{\cal D}X^A \exp\left[-S_{\rm B}-\int_0^\beta d\tau H(\tau)\right],
\end{equation}
where $S_{\rm B}$ represents the Berry phase term, whose explicit form is irrelevant to discussions to follow.
We use the Stratonovich-Hubbard identity for each imaginary time interval, then the direct interaction between multipoles are replaced by indirect ones via the time-dependent (molecular) fields $\phi^A(\tau)$, which are conjugate to $X^A(\tau)$:
\begin{equation}
Z=\int{\cal D}X^A{\cal D}\phi^A \exp\left[-S_{\rm B}-\int_0^\beta d\tau H_\phi(\tau)\right],
\end{equation}
where $H_\phi(\tau)$ is given by
\begin{equation}
H_\phi=\frac{1}{2}\sum_{ij}\left({\sf D}^{-1}\right)^{AB}_{ij}\phi^A_i\phi^B_j-\sum_i\phi^A_iX^A_i.
\end{equation}
Making the static approximation for the fluctuating field $\phi^A$, we obtain the semi-classical form of the partition function as,
\begin{equation}
Z=\int {\cal D}\phi^A \exp\left[-\beta{\cal F}\right],
\label{part}
\end{equation}
where the GL free energy functional is given by
\begin{eqnarray}
&&{\cal F}=-NT\ln 4+\frac{1}{2}\sum_{ij}\left[({\sf D}^{-1})^{AB}_{ij}-\beta\delta_{AB}\delta_{ij}\right]\phi^A_i\phi^B_j\nonumber\\
&&\mbox{\hspace{1cm}}+Nf_{\rm mc},
\end{eqnarray}
with
\begin{equation}
f_{\rm mc}=-\frac{1}{\beta N}\sum_i\left[\ln\left({\rm Tr}_i\exp(\beta\phi^A_i X^A_i)/4\right)-\frac{\beta^2}{2}\phi^A_i\phi^A_i\right],
\label{free-e}
\end{equation}
where $N$ is the number of sites, and the trace is carried out at site $i$.
The first term in ${\cal F}$ is the entropy of the non-interacting system, and $f_{\rm mc}$ describes the mode coupling between multipole fluctuations beyond the RPA.
Note that the non-commuting character of quantum multipoles is kept faithfully in this approximation scheme due to the exact trace for the quantum operators $X^A_i$.
Hence we expect that interesting consequences due to coupling between different multipoles can be understood qualitatively within the static approximation.
It is obvious from eq. (\ref{free-e}) that the mode coupling among different multipoles comes from purely local effect.
Note that the pseudo-dipole interactions mix the different modes $\phi^A$ and $\phi^B$ even in the Gaussian term.

The generalized susceptibilities for multipole moments and conjugate fields are defined as the second-order cumulants:
\begin{eqnarray}
&&
\chi^{AB}_{ij}=\beta\left(\langle X^A_iX^B_j\rangle-\langle X^A_i\rangle\langle X^B_j\rangle\right),
\\&&
G^{AB}_{ij}=\beta\left(\langle \phi^A_i\phi^B_j\rangle-\langle \phi^A_i\rangle\langle \phi^B_j\rangle\right).
\end{eqnarray}
Note that with the Gaussian distribution of $\phi^A$, $G^{AB}_{ij}$ is given by its variance and physically means the renormalized interaction between the multipoles.
These quantities are related to each other in the matrix representation as
\begin{equation}
{\sf \chi}={\sf D}^{-1}{\sf G} {\sf D}^{-1}-{\sf D}^{-1}.
\label{relchig}
\end{equation}
Moreover the order parameter and the symmetry-breaking field are connected by the linear combination,
\begin{equation}
\langle{\mib \phi}\rangle={\sf D}\langle {\mib X}\rangle.
\label{oprel}
\end{equation}
In what follows, we apply the saddle point approximation to eq. (\ref{part}).
Hence, resultant solutions are equivalent to those obtained by the mean-field approximation.
The present formalism gives much simpler calculational scheme than the standard mean-field theory.

Let us first determine the quadrupolar instability within the RPA by neglecting the mode coupling term $f_{\rm mc}$.
The critical temperature is determined by $|{\sf \chi}^{-1}|=0$.
This condition is equivalent to $|{\sf D}{\sf G}^{-1}|=|1-\beta{\sf D}|=0$ due to the relation (\ref{relchig}).

The pseudo-dipole interaction of eq.(\ref{pdint}) has the maximum value at ${\mib q}=(1/4,1/4,1/4)$, while the nearest-neighbor interaction of eq. (\ref{nnint}) at $(1/2,1/2,1/2)$.
Thus, the wave vector giving the maximum eigenvalue of ${\sf D}_{\Gamma}$ becomes incommensurate for large $|K_{\Gamma}/J_{\Gamma}|$, i.e., unless $-1/2 \le K_{\Gamma n}/J_{\Gamma n} \le 1$.
The incommensurate vector $q^*(111)$ is given by $q^*=\cos^{-1}(-1/s)$ and the corresponding maximum eigenvalue is
\begin{equation}
T_c^{(\Gamma)}\equiv D_{\Gamma}^{\rm max}=\frac{J_{\Gamma}}{2}(s+s^{-1}),
\end{equation}
where
\begin{equation}
s=\left\{\begin{array}{ll}
K_{\Gamma}/J_{\Gamma} & (K_{\Gamma}/J_{\Gamma}>1), \\
2|K_{\Gamma}|/J_{\Gamma} & (K_{\Gamma}/J_{\Gamma} < -1/2).
\end{array}\right.
\end{equation}
Otherwise the AF wave vector ${\mib Q}\equiv(1/2,1/2,1/2)$ gives the maximum eigenvalue.
The stable condition for the AF quadrupole order of the $\Gamma_{5g}$-type is given by
\begin{equation}
T_Q\equiv J_{5g}>\max(T_c^{(\Gamma)},J_{\Gamma})\;\;\;(\Gamma\ne 5g).
\label{tqc}
\end{equation}
The possibility of the incommensurate ordering will be discussed later.

The wave vector ${\mib q}$ is equivalent to ${\mib q}+{\mib Q}$ under the AF quadrupolar ordering.
In addition, certain multipole modes are mixed due to a lowering of the point-group symmetry.
In what follows we work with the single domain characterized by the $O_{xy}$-component quadrupole.
To determine the lower critical temperature under the AF $O_{xy}$ ordering, we collect quadratic terms for fluctuation in the mode-coupling free energy (\ref{free-e}), and consider faithfully the finite order parameter $\xi\equiv\langle \phi^{5gz}_{\mibs Q}\rangle/T_Q\ne 0$.
Using the perturbation expansion (see Appendix A for detail), we obtain the GL free energy under the $O_{xy}$ ordering:
\begin{equation}
f_{\rm mc}=f_a+f_b+f_c,
\label{foxy}
\end{equation}
with
\begin{eqnarray}
&&f_a=\frac{1}{2}\sum_{\mibs q}\biggl[
(\beta-T_Q^{-1}){\mib\phi}^{'4u1}_{\mibs q}\cdot{\mib\phi}^{'4u1}_{-{\mibs q}}
-\beta\xi\phi^{2u}_{\mibs q}\cdot\phi^{4u1z}_{-({\mibs q}+{\mibs Q})}
\biggr],\nonumber\\ \\
&&f_b=\frac{1}{2}\sum_{\mibs q}\biggl[
(\beta-T_Q^{-1})(\phi^{4u2z}_{\mibs q}\cdot\phi^{4u2z}_{-{\mibs q}}+\phi^{5uz}_{\mibs q}\cdot\phi^{5uz}_{-{\mibs q}})\nonumber\\&&\mbox{\hspace{1cm}}
+\frac{1}{2}\beta\xi\biggl\{
\sqrt{3}(\phi^{4u2x}_{\mibs q}\cdot\phi^{4u2y}_{-({\mibs q}+{\mibs Q})}+\phi^{5ux}_{\mibs q}\cdot\phi^{5uy}_{-({\mibs q}+{\mibs Q})})\nonumber\\
&&\mbox{\hspace{1cm}}
+(\phi^{4u2y}_{\mibs q}\cdot\phi^{5ux}_{-({\mibs q}+{\mibs Q})}-\phi^{4u2x}_{\mibs q}\cdot\phi^{5uy}_{-({\mibs q}+{\mibs Q})})
\biggr\}
\biggr], \\
&&f_c=\frac{1}{2}\sum_{\mibs q}\biggl[
(\beta-T_Q^{-1})\bigl({\mib\phi}^{3g}_{\mibs q}\cdot{\mib\phi}^{3g}_{-{\mibs q}}+{\mib\phi}^{'5g}_{\mibs q}\cdot{\mib\phi}^{'5g}_{-{\mibs q}}\bigr)\nonumber\\&&\mbox{\hspace{1cm}}
+\beta\xi^2\delta\phi^{5gz}_{\mibs q}\cdot\delta\phi^{5gz}_{-{\mibs q}}
\biggr],
\label{freec}
\end{eqnarray}
where ${\mib\phi}^{3g}$ is a two-component vector, ${\mib\phi}^{'\Gamma}=(\phi^{\Gamma x},\phi^{\Gamma y})$ and $\delta\phi^{5gz}_{\mibs q}=\phi^{5gz}_{\mibs q}-\langle\phi^{5gz}_{\mibs Q}\rangle\delta_{\mibs q,Q}$.
The average $\langle\phi^{5gz}_{\mibs Q}\rangle$ satisfies the following mean-field equation\cite{mfnote},
\begin{equation}
\xi\equiv \frac{\langle \phi^{5gz}_{\mibs Q}\rangle}{T_Q}=\tanh(\beta\langle\phi^{5gz}_{\mibs Q}\rangle).
\end{equation}
In the above expressions, the mode-mixing terms among ($2u$, $4u1z$) and ($4u2x$, $4u2y$, $5ux$, $5uy$) are proportional to $\beta\xi$.

The part $f_c$ gives no further instability as long as the condition (\ref{tqc}) is satisfied.
Namely, $f_c$ and the corresponding Gaussian terms give temperature-independent ``transverse" susceptibility for ${\mib X}^{3g}$ and ${\mib X}^{'5g}$ components and ``longitudinal" one for the other component as is similar to the case of the isotropic Heisenberg model.
The lower transition temperature $T_N$ will be determined by $|{\sf \chi}^{-1}|=0$ for each part, $f_a$ or $f_b$.
The normal mode of coupled order parameters is determined by the eigenvalue equation, $({\sf \chi}^{-1}){\mib \phi}=0$ at $T=T_N$.

\section{Mean-Field Solutions}
\subsection{Limit of $T_Q/T_N\gg 1$}
If interactions corresponding to the lower magnetic phase transition are weak, the limit $T_N^{-1}\gg \beta\xi\gg T_Q^{-1}$ is realized.
The effective Hamiltonian in this limit is given in Appendix B.
Let us discuss the lower transition with the strong mode mixing.
In the limit of $\beta\xi\to\infty$, the local free energies are expressed in terms of normal modes:
\begin{eqnarray}
&&f_{ai} \sim \beta\xi_i[(\phi^{a+}_i)^2-(\phi^{a-}_i)^2],\\
&&f_{bi} \sim \beta\xi_i[(\phi^{b+}_i)^2+(\varphi^{b+}_i)^2-(\phi^{b-}_i)^2-(\varphi^{b-}_i)^2],\label{fcq}
\end{eqnarray}
where the normal modes have been defined as
\begin{equation}
\phi^{a\pm}_i=\frac{1}{\sqrt{2}}(\phi_i^{2u}\mp \phi^{4u1z}_i),
\label{normmodeb}
\end{equation}
in $f_a$ and
\begin{eqnarray}
&&\phi^{b\pm}_i=\frac{1}{2}[A_-(\phi^{4u2x}_i\mp \phi^{4u2y}_i)-A_+(\phi^{5ux}_i\pm \phi^{5uy}_i)],\nonumber\\
&&\varphi^{b\pm}_i=\frac{1}{2}[A_+(\phi^{4u2x}_i\pm \phi^{4u2y}_i)+A_-(\phi^{5ux}_i\mp \phi^{5uy}_i)],\nonumber\\
\label{normmodec}
\end{eqnarray}
in $f_b$.
Here we have defined $A_\pm=(\sqrt{3}\pm1)/2$.
The normal modes $\phi^{a\pm}_i$, $\phi^{b\pm}_i$ and $\varphi^{b\pm}_i$ give the lowest energy for the sub-lattice $\xi_i=\mp$.

\begin{figure}
\begin{center}
\epsfxsize=8cm \epsfbox{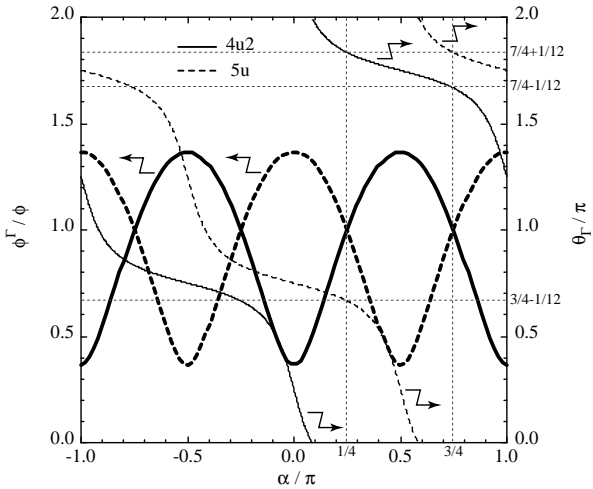}
\end{center}
\caption{The molecular fields at sublattice $\xi_i=+$, ${\mib\phi}^\Gamma\equiv\phi^\Gamma(\cos\theta_\Gamma,\sin\theta_\Gamma,0)$, in the limit of $T_Q/T_N\gg 1$. The solid (dotted) line represents the case of $\Gamma=4u2$ ($\Gamma=5u$). The free energy is independent of $\alpha$.}
\label{normalpha}
\end{figure}

In $f_b$ the normal modes are doubly degenerate.
Thus, any linear combination of degenerate modes gives the same free energy and the degeneracy is lifted only by intersite interactions.
If we set $(\phi^{b-}_i,\varphi^{b-}_i)=\sqrt{2}\phi(\cos\alpha,\sin\alpha)$ for sublattice $\xi_i=+$, the local molecular fields depend on $\alpha$ as shown in Fig. \ref{normalpha}, in which $(\phi^{\Gamma x}_i,\phi^{\Gamma y}_i)=\phi^\Gamma(\cos\theta_\Gamma,\sin\theta_\Gamma)$.
It is emphasized that the easy axis of the $\Gamma_{4u2}$ dipole moment corresponds to the hard axis of the $\Gamma_{5u}$ octupole moment ($\alpha=\pm\pi/2$) and vice versa ($\alpha=0,\pi$).
Therefore, when the $\Gamma_{4u2}$-type dipole-dipole and the $\Gamma_{5u}$-type octupole-octupole interactions are comparable, the ground state characterized by $\alpha=\pm\pi/2$ and the excited states by $\alpha=0,\pi$ are almost degenerate.
In such a situation, weak perturbations easily change one ground state to the other.

\subsection{Case of the nearest-neighbor interactions}
\begin{figure}
\begin{center}
\epsfxsize=16cm \epsfbox{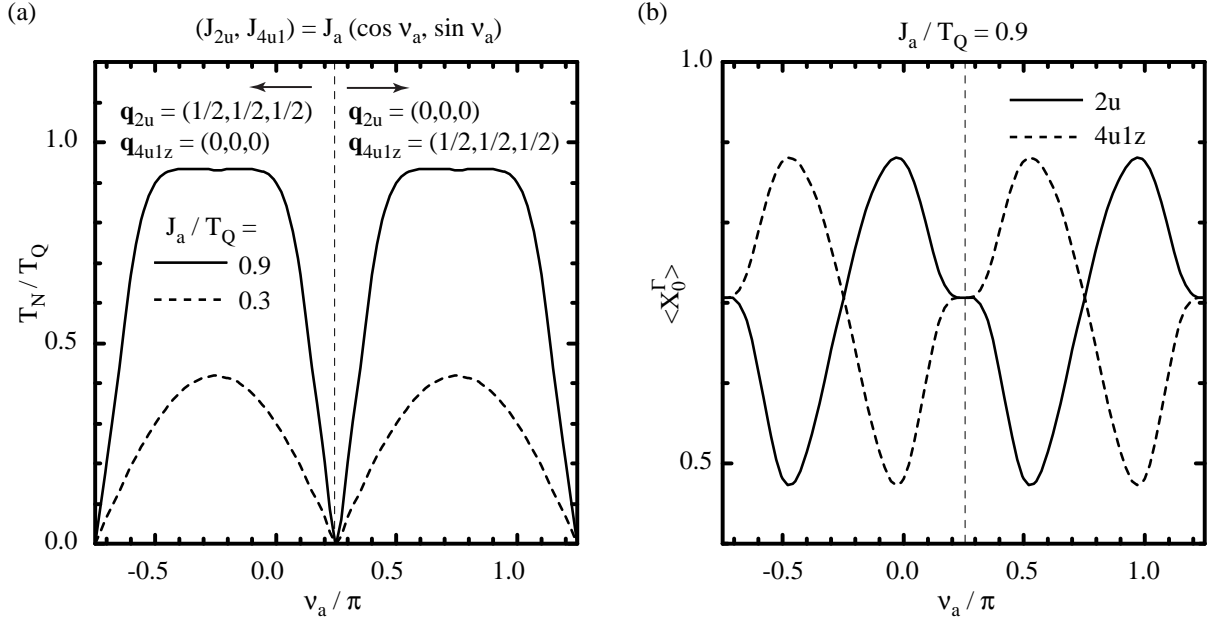}
\end{center}
\caption{The lower transition temperature (a), and the order parameters at the origin (b) in the absence of the next-neighbor interactions.
The phase of the coexistent AF (F) octupole and the F (AF) dipole is realized for $\nu_a<\pi/4$ ($\nu_a>\pi/4$).}
\label{fbnn}
\end{figure}

Let us now investigate the case of the nearest neighbor interactions only, i.e., $K_{\Gamma}=0$.
First, we discuss the phase transition determined by $f_a$ for interactions parameterized by $(J_{2u},J_{4u1})=J_a(\cos\nu_a,\sin\nu_a)$.
The order parameters have been normalized as $\langle X^{2u}_i \rangle^2+\langle X^{4u1z}_i\rangle^2=1$.
As shown in Fig. \ref{fbnn}, the AF octupole order $\langle X^{2u}_{\mibs Q}\rangle$ with the ferro dipole order $\langle X^{4u1z}_{\mibs 0}\rangle$ occurs for $\nu_a < \pi/4$, while for $\nu_a>\pi/4$ ordered vectors for the octupole and the dipole is interchanged.
When the $\Gamma_{4u1}$-type dipole-dipole and the $\Gamma_{2u}$-type octupole-octupole interactions are comparable with the same sign, the competition between them suppresses $T_N$, while the opposite sign of interactions work cooperatively to stabilize the coexistent phases\cite{sera99,shiba99}.
As $J_a$ decreases, the system can be described by the limit of $T_Q/T_N\gg 1$ as was discussed in the previous subsection.
Therefore the order parameter becomes almost independent of $\nu_a$ and it approaches to $\phi^{a-}_i$ of eq. (\ref{normmodeb}) for ${\mib R}_i={\mib 0}$.

\begin{figure}
\begin{center}
\epsfxsize=16cm \epsfbox{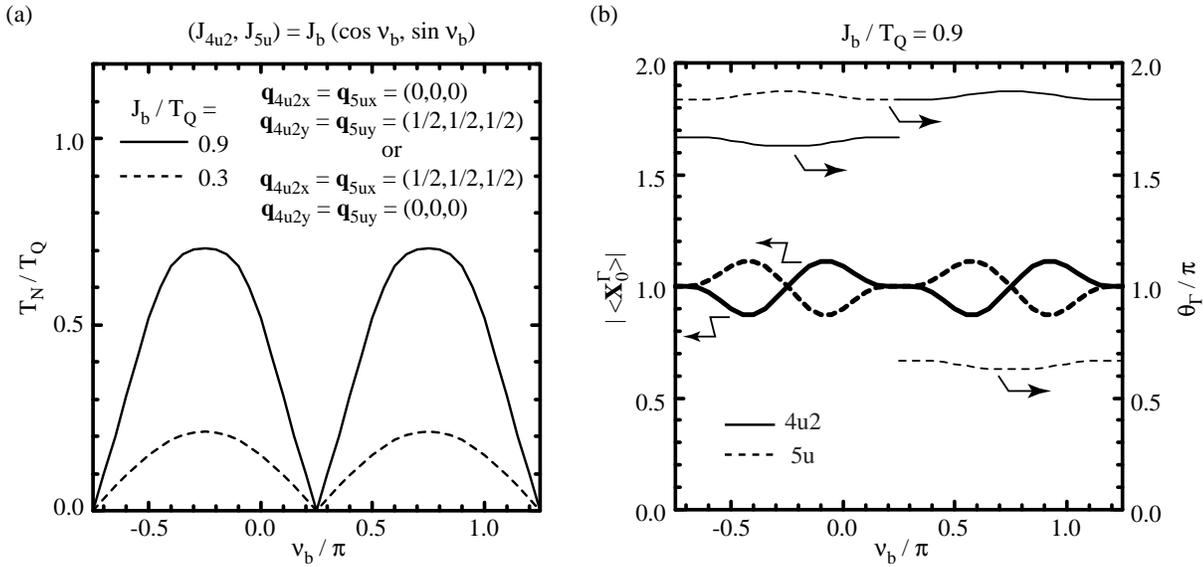}
\end{center}
\caption{The lower transition temperature (a), and the order parameters at the origin, $\langle{\mib X}^\Gamma_0\rangle\equiv|\langle{\mib X}^\Gamma_0\rangle|(\cos\theta_\Gamma,\sin\theta_\Gamma,0)$, (b) in the absence of the next-neighbor interactions.
The canted ordering with the modulation vector $(0,0,0)$ or $(1/2,1/2,1/2)$ is realized because of the coexistence of $J_{4u2}$ and $J_{5u}$.}
\label{fcnn}
\end{figure}

Next, the corresponding results for the phase transition determined by $f_b$ are shown in Fig. \ref{fcnn} for interactions parameterized by $(J_{4u2},J_{5u})=J_b(\cos\nu_b,\sin\nu_b)$.
We have used the normalization, $\langle{\mib X}^{4u2}_i\rangle^2+\langle{\mib X}^{5u}_i\rangle^2=2$.
As is similar to the previous case, the competition between $J_{4u2}$ and $J_{5u}$ causes to decrease $T_N$.
This is because the $\Gamma_{4u2}$-type dipole and the $\Gamma_{5u}$-type octupole cannot point simultaneously to the common easy-axis along $[1,\mp1,0]$ direction for sublattice $\xi_i=\pm$.
As the difference between $J_{4u2}$ and $J_{5u}$ increases, the phase of coexistent $\langle X^{4u2}_{\mibs q}\rangle$ and $\langle X^{5u}_{{\mibs q}+{\mibs Q}} \rangle$ is stabilized at ${\mib q}={\mib 0}$ or ${\mib Q}$.
In order to gain the nearest neighbor interactions while keeping the same magnitude of $\langle{\mib X}^{4u2}\rangle$ and $\langle{\mib X}^{5u}\rangle$, the moments are rotated about $\pi/12$ radian from the easy axis as shown in Fig. \ref{fcnn}(b).
Namely, setting $\langle {\mib X}^\Gamma_0\rangle=|\langle{\mib X}^\Gamma_0\rangle|(\cos\theta_\Gamma,\sin\theta_\Gamma,0)$, we obtain $(\theta_{4u2},\theta_{5u})/\pi\sim(7/4,7/4)+(-1/12,1/12)$ for $\nu_b<\pi/4$ and $\sim(7/4,3/4)+(1/12,-1/12)$ for $\nu_b>\pi/4$.
As $J_b$ decreases, the order parameter approaches to the linear combination of $\phi^{b-}_i$ and $\varphi^{b-}_i$ of eq.~(\ref{normmodec}) with $\alpha=3\pi/4$ ($\alpha=\pi/4$) for $\nu<\pi/4$ ($\nu>\pi/4$) (see Fig. \ref{normalpha}).

The phases discussed above have not been observed in CeB$_6$ so far.
This fact indicates that there are strong competitions between the nearest neighbor interactions of $J_{2u}$ and $J_{4u1}$, and of $J_{4u2}$ and $J_{5u}$\cite{shiba99}.

\subsection{Case of the next-nearest-neighbor interactions}
\begin{figure}
\begin{center}
\epsfxsize=16cm \epsfbox{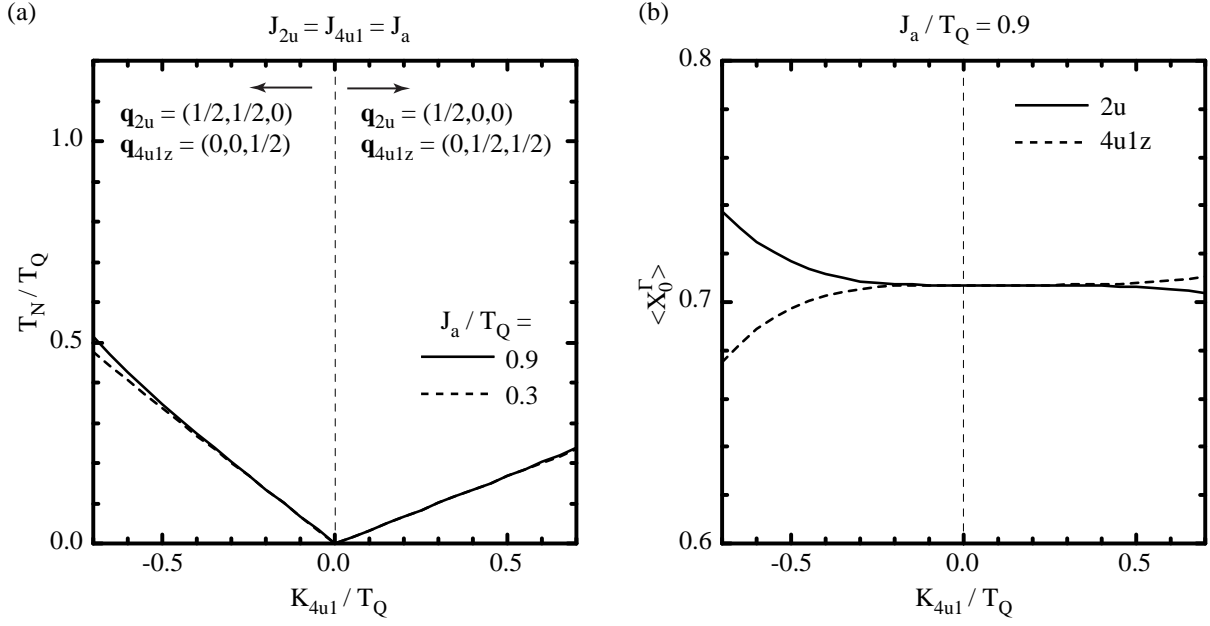}
\end{center}
\caption{The lower transition temperature (a) and the order parameters at the origin (b). The type I AF $(0,0,1/2)$ dipole ordering with the $(1/2,1/2,0)$ octupole one is realized for $K_{4u1}<0$.
The type I octupole ordering with the $(0,1/2,1/2)$ dipole one is realized for $K_{4u1}>0$.}
\label{fbnnn}
\end{figure}

Let us now take into account the next-nearest neighbor interactions.
We restrict ourselves to the case of the perfect competition between the nearest neighbor interactions, i.e., $J_{2u}=J_{4u1}$ and $J_{4u2}=J_{5u}$.
Figure \ref{fbnnn} shows $T_N$ and the order parameters determined by $f_a$.
The ordering vector for the $\Gamma_{4u1}$-type dipole moment is given by ${\mib Q}_z=(0,0,1/2)$ for $K_{4u1}<0$, while by ${\mib Q}_{yz}=(0,1/2,1/2)$ for $K_{4u1}>0$.
Note that the ordering vector for the $\Gamma_{2u}$-type octupole is given by ${\mib Q}_n-{\mib Q}$ ($n=z$ or $yz$).
Interestingly, the type I magnetic order has been observed in NdB$_6$\cite{loewenhaupt86,pofahl87,erkelens88}, which has a $\Gamma_8$ CEF ground state as in CeB$_6$.

\begin{figure}
\begin{center}
\epsfxsize=16cm \epsfbox{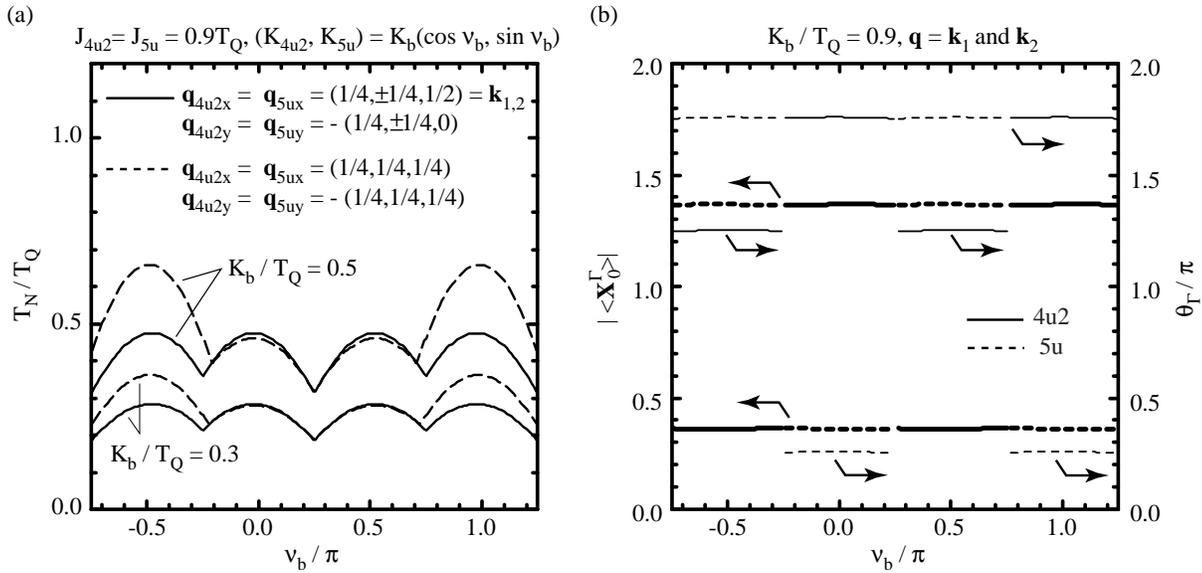}
\end{center}
\caption{The lower transition temperature determined by the instability in $f_b$. The observed pattern of the non-collinear magnetic ordering is realized for $-\pi/4\lsim\nu_b\lsim\pi/4$. The order parameters with the double ${\mib k}$ structure are shown in (b).}
\label{fcnnn}
\end{figure}

The observed non-collinear ordering of phase III arises from the instability of $f_b$.
The instability will occur at ${\mib q}={\mib Q}^\prime\equiv(1/4,1/4,1/4)$ or ${\mib k}_i$ ($i=1,2$) because $D_{\Gamma}$ for these wave vectors has the same maximum eigenvalue for positive $K_{\Gamma}$.
As shown in Fig. \ref{fcnnn}, the magnetic ordering observed in phase III is indeed realized for $-\pi/4 \lsim \nu_b \lsim \pi/4$, where $(K_{4u2},K_{5u})=K_b(\cos\nu_b,\sin\nu_b)$ with $J_{4u2}=J_{5u}=0.9T_Q$.
Note that the phase given by $\pi/4<\nu_b<5\pi/4$ has a ordering pattern different from the observed one.
The transition temperatures for ${\mib Q}^\prime$ and ${\mib k}_1$ (${\mib k}_2$) become the same as $T_N$ (or $K_b$) decreases.
This is because the effective Hamiltonian for $\beta\xi\gg T_Q^{-1}$ has the same form for these wave vectors (see, in Appendix B).
In practice, however, the transition temperatures may be different even in this limit due to the anisotropy in the interactions brought about by the quadrupolar ordering.

\begin{figure}
\begin{center}
\epsfxsize=16cm \epsfbox{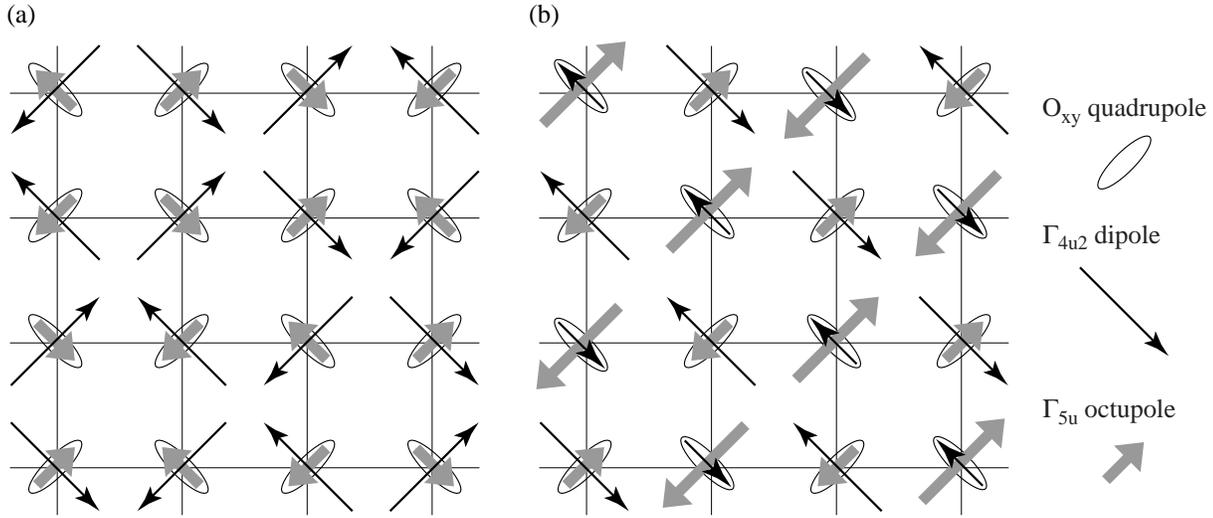}
\end{center}
\caption{The ordering pattern for dipole and octupole moments: (a) the mean-field solution for $-\pi/4<\nu_b<\pi/4$ corresponding to the phase III, (b) the phase III' under the magnetic field along $[111]$. The oval represents $O_{xy}$ quadrupole, thin and thick arrows $\Gamma_{4u2}$-type dipole and $\Gamma_{5u}$-type octupole moments, respectively.}
\label{spinpat}
\end{figure}

As for the real-space ordering pattern, the combination of ${\mib k}_1$ and ${\mib k}_2$ is consistent with the AFQ ordering in the absence of the external magnetic field.
By choosing arbitrary phases such that the order parameters at ${\mib R}_i=(1,0,0)$ are equal to those at ${\mib R}_i=(0,0,1)$, the orientation of the order parameters for $-3\pi/4<\nu_b<\pi/4$ is expressed in terms of $\langle X^A_i\rangle$ at the origin, ${\mib R}_i=(0,0,0)$, as shown in Fig. \ref{fcnnn}(b):
\begin{eqnarray}
\langle {\mib X}^{4u2}_i \rangle &=&
{\mib u}_{k1}\left[\langle X^{4u2x}_0 \rangle c_1-\langle X^{4u2y}_0 \rangle c_1^\prime\right]\nonumber\\
&+&{\mib u}_{k2}\left[\langle X^{4u2x}_0 \rangle c_2+\langle X^{4u2y}_0 \rangle c_2^\prime\right],\\
\langle {\mib X}^{5u}_i \rangle &=&
{\mib u}_{k1}\left[\langle X^{5ux}_0 \rangle c_2-\langle X^{5uy}_0 \rangle c_2^\prime\right]\nonumber\\
&+&{\mib u}_{k2}\left[\langle X^{5ux}_0 \rangle c_1+\langle X^{5uy}_0 \rangle c_1^\prime\right],
\end{eqnarray}
where we have used the unit polarization vectors, ${\mib u}_{k1}=(1,-1,0)/\sqrt{2}$ and ${\mib u}_{k2}=(110)/\sqrt{2}$, and $c_n=\cos({\mib k}_n\cdot{\mib R}_i+\pi/4)$ and $c_n^\prime=\cos({\mib k}_n^\prime\cdot{\mib R}_i-\pi/4)$ with ${\mib k}_n^\prime={\mib k}_n+{\mib Q}$ for $n=1,2$.
For $\pi/4<\nu_b<5\pi/4$, the indices 1 and 2 for $n$ are interchanged in the above expressions.
The ordering pattern for $-\pi/4<\nu_b<\pi/4$ in the plane $z=0$ is shown in Fig. \ref{spinpat}(a), where the oval represents $O_{xy}$ quadrupole, and thin (thick) arrow represents $\Gamma_{4u2}$-type dipole ($\Gamma_{5u}$-type octupole) moment.
For $\nu_b<-\pi/4$ the magnitude of the octupole moment is larger than the dipole one because of the stronger octupole-octupole interaction, although the ordering pattern is the same as the case $-\pi/4<\nu_b<\pi/4$.

For a magnetic field applied along $[111]$, the phase III$^\prime$ characterized by the single ${\mib k}_1$ appears experimentally at $H_c=10.5$ kOe and $T=1.5$ K\cite{effantin85,erkelens87}.
As mentioned in \S3.1, when $K_{4u2}$ and $K_{5u}$ are comparable, excited states with $|{\mib X}^{5u}|$ larger than $|{\mib X}^{4u2}|$ are energetically close to the ground state.
The fact that rather small magnetic field drives the phase transition to the phase III$^\prime$ indicates $K_{4u2}\sim -K_{5u}$ ($\nu_b\sim -\pi/4$).
In other words, the energy loss by shortening the dipole moment can be compensated by the extension of the octupole moment as shown in Fig. \ref{spinpat}(b).
Indeed, the observed spin pattern in phase III$^\prime$ is constructed by the combination of the patterns obtained in cases $-3\pi/4<\nu_b<-\pi/4$ and $-\pi/4<\nu_b<\pi/4$.
The ratio observed ($0.26/0.08\sim3.25$) of two distinct moments\cite{effantin85} is quite close to $A_+/A_-\sim 3.73$.

\begin{figure}
\begin{center}
\epsfxsize=8cm \epsfbox{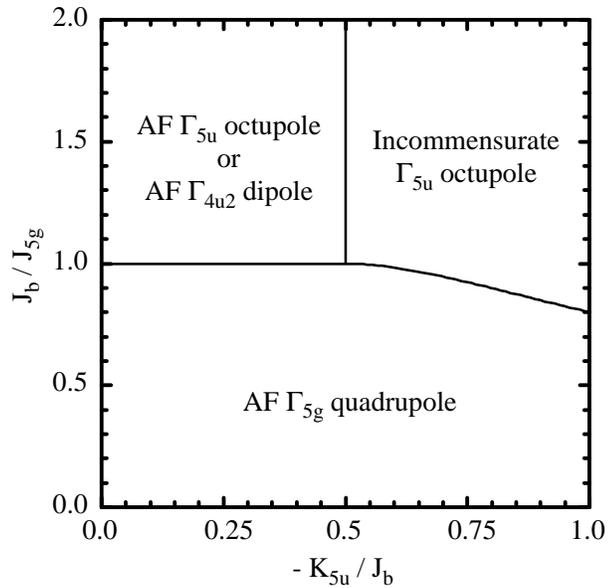}
\end{center}
\caption{Parameter dependence of the most stable upper ordered phases for $J_b\equiv J_{4u2}=J_{5u}$ and $1/2<-K_{5u}/K_{4u2}<1$.}
\label{octphase}
\end{figure}

Putting together the above analysis, we consider that the parameters for CeB$_6$ are located near the boundary given by the stable condition (\ref{tqc}) of $\Gamma_{5g}$-type quadrupolar phase.
In order to discuss the alternative phase transition, which may happen with the slight modification of interactions, the parameter dependence of the most stable upper ordered phases are shown in Fig. \ref{octphase}, where $J_b=J_{4u2}=J_{5u}$ and $1/2<-K_{5u}/K_{4u2}$.
If we use $J_b/T_Q=0.9$ ($\sim J_{5g}$) and $-K_{5u}/K_{4u2}=0.9$, then $-K_{5u}/J=K_{4u2}/T_Q\sim T_N/T_Q\sim 0.68$.
The parameters, $(J_b/J_{5g},-K_{5u}/J_b)\sim (0.9,0.68)$ is located close to the stable domain of the incommensurate $\Gamma_{5u}$-type octupole phase at ${\mib q}=q^*(111)$.

Assuming the incommensurate $\Gamma_{5u}$-type octupole ordering, we obtain the Legendre transform of the free energy $G(\langle X^A\rangle)$ as (see Appendix C)
\begin{eqnarray}
&&G/T_c^{5u}=\frac{1}{2}(t-1)|\bar{X}^{5u}_{q^*}|^2 +\frac{1}{2}(t+\frac{T_Q}{T_c^{(5u)}})(\bar{X}^{5g}_{\mibs 0})^2
+\frac{t}{2}|\bar{X}^{5u}_{q^*}|^2\bar{X}^{5g}_{\mibs 0}
\nonumber\\&&\mbox{\hspace{8mm}}
+\frac{t}{12}\bigl[2|\bar{X}^{5u}_{q^*}|^4+(\bar{X}^{5g}_{\mibs 0})^4\bigr]+\cdots,
\end{eqnarray}
where
\begin{equation}
\bar{X}^{\Gamma}_{\mibs q}=\frac{1}{\sqrt{3}}\bigl\langle X^{\Gamma x}_{\mibs q}+X^{\Gamma y}_{\mibs q}+X^{\Gamma z}_{\mibs q}\bigr\rangle,
\end{equation}
and $t\equiv T/T_c^{(5u)}$.
Here the higher harmonics ($2q^*$, $3q^*$, etc.) of the order parameters have not been written explicitly since they do not influence the temperature dependence of the order parameters.
For $T_c^{(5u)}/T_Q<2$, the transition is of second order.
Thus, in the presence of the $\bar{X}^{5u}_{q^*}$ order, the ferro-quadrupole order of the $\Gamma_{5g}$-type appears as the secondary order parameter.
The temperature dependence of the order parameters is given by $\bar{X}^{5g}_{\mibs 0}\propto|\bar{X}^{5u}_{q^*}|^2\propto(1-t)$.
Note that the susceptibility of the secondary order parameter does not show critical divergence.
Recent neutron diffraction study has revealed anomalous increases of nuclear Bragg-reflection intensities in the phase IV\cite{iwasa00}.
However, no lattice deformations have been observed so far.
At present it is not certain whether $\bar{X}^{5g}_{\mibs 0}$ is absent or too small to be observed.

In the discussion above, we have neglected possible orderings of $\Gamma_{2u}$\cite{kuramoto00}, $\Gamma_{4u1}$ and $\Gamma_{3g}$ types.
Within the present model calculation, such orderings are not related to the existence of the phase III and are realized near $T_Q$ only by accidental tuning of interactions.
However, if there are interactions neglected in the present model, such as the mixing between $\Gamma_{4u1}$ and $\Gamma_{4u2}$, then $f_a$ and $f_b$ are coupled with each other and consequently with the interactions leading to the phase III.
We note that the pure $\Gamma_{2u}$-type octupole ordering is also discussed for NpO$_2$\cite{santini00}.

%
%
\section{Summary}
We have investigated the magnetic phase transition under the AF quadrupolar ordering keeping CeB$_6$ in mind.
The GL free energy has been classified microscopically by the systematic analysis of the mode mixing among the 15 multipoles.
We have determined the lower transition temperature and the order parameters from the instability of each part of the free energy.

In the presence of the pseudo-dipole-type interactions for the next-nearest neighbor dipole and octupole moments, the non-collinear ordering observed in the phase III is indeed stabilized for the positive coupling of the dipole-dipole interaction.
The negative octupole-octupole interaction is expected to be equally strong since the phase III$^\prime$ is stabilized with rather weak magnetic field.
The strength of interactions for CeB$_6$ seem to be located close to the boundary of the stable condition for the AF quadrupolar phase.
This indicate that the slight modification of interactions stabilizes the incommensurate $\Gamma_{5u}$-type octupole ordering, where the $\Gamma_{5g}$-type ferro-quadrupole order parameter appears as the secondary one.
We note that the magnitude $q^*$ of the incommensurate order should be influenced by the presence of other interactions neglected in this paper.
Thus experimental and theoretical investigations to derive actual intersite interactions are desired in the near future.

At zero temperature the saturated moment becomes $|{\mib M}|=4/7\cdot A_+\sim 0.78\mu_{\rm B}$ in the present theory, which is much larger than the observed value, $0.28\mu_{\rm B}$\cite{effantin85,erkelens87}.
Thus, careful characterization of quantum fluctuations should be important.
If the quantum fluctuation is significant, a characteristic energy scale of magnetic fluctuation should be larger\cite{nakamura00} than the widely believed value of 1K\cite{sato85}.
The same magnetic pattern as that in the phase III has been observed in PrB$_6$\cite{mccarthy80,loewenhaupt86,burlet88}, in which there is no octupole moments within the $\Gamma_5$ CEF ground state.
The comparative study between both compounds will be rewarding in future investigation.

%
%
\section*{Acknowledgements}
H. K. would like to thank O. Sakai for fruitful discussion and suggestion of using pseudo-dipole interactions.
He has also benefited from stimulating conversations with N. Fukushima, T. Kuromaru, S. Nakamura and T. Matsumura.

%
%
\appendix
\section{Derivation of Mode-Coupling Free Energy under AF Ordering}
Here, we give details of the perturbation expansion for the mode-coupling free energy, eq. (\ref{free-e}) in the presence of an AF order parameter $\langle X^0_{\mibs Q} \rangle\ne0$, where we set $A=0$ as the index of the order parameter.
Let us denote the unperturbative part by $H_0=-\phi^0_iX^0_i$ and the rest by $H_1=-\sum_A'\phi^A_iX^A_i$, where the prime means that the summation for $A$ runs from 1 to 15 except for a certain value corresponding to $A=0$.
Using the standard perturbation expansion, we obtain the following form up to the second-order in $H_1$:
\begin{eqnarray}
&&
\ln{\rm Tr}_i\left(e^{-\beta(H_0+H_1)}\right)=\ln Z_0+S_1+S_2,\\
&&
S_1=-\beta\langle H_1 \rangle_0,\;\;
S_2=\frac{1}{2}\biggl[\beta^2 \sum_{AB}{}^\prime \Pi^{AB}_i \phi^A_i\phi^B_i-S_1^2\biggr],\\
&&
\Pi^{AB}_i=
\nonumber\\&&\mbox{\hspace{5mm}}
 \frac{1}{\beta^2} \int_0^\beta d\lambda \int_0^\lambda d\xi \biggl\langle e^{\xi H_0} X^A_i e^{-\xi H_0} X^B_i + (A \leftrightarrow B)\biggr\rangle_0, \nonumber\\
\end{eqnarray}
where the average is defined by $\langle A\rangle_0={\rm Tr}_i(e^{-\beta H_0}A)/Z_0$ with $Z_0={\rm Tr}_ie^{-\beta H_0}$.

The following formula for the product of tensor operators is useful:
\begin{equation}
X^AX^B=\left(if_{ABC}+g_{ABC}\right)X^C+\delta^{AB}I,
\end{equation}
where $f_{ABC}$ is the antisymmetric structure constant.
Note that the symmetric constant $g_{ABC}$ appears here in contrast to the case of two-fold degeneracy, in which only the two-dimensional Pauli matrices are involved and thus $g_{ABC}$ vanishes.
Both constants in the case of the active multipoles in the $\Gamma_8$ state are tabulated in Tables II and III.

If the operator in $H_0$ has the property, $(X^0_i)^2=I$, as is the case with all operators listed in Table I, the exponential factor can be reduced to
\begin{equation}
\exp(\pm\xi H_0) = \cosh(\xi\phi^0_i)I\mp \sinh(\xi\phi^0_i)X^0_i,
\end{equation}
we obtain $Z_0=4\cosh(\beta\phi^0_i)$ and $S_1=0$.
We insert the above expression into $\Pi^{AB}_i$ and use the following relations,
\begin{eqnarray}
&&
(ff)^0_{AB}=(gg)^0_{AB}-\delta^{AB}+\delta^{A0}\delta^{B0},\nonumber \\
&&
(gf)^0_{AB}=(fg)^0_{AB}=0,\\
&&
{\rm Tr}_i ( X^A_iX^B_i)/4 = \delta^{AB}, \nonumber \\
&&
{\rm Tr}_i (X^0_i\{X^A_i,X^B_i\})/4 =2g_{0AB}, \nonumber \\
&&
{\rm Tr}_i (X^0_iX^A_iX^0_iX^B_i)/4 = 2(gg)^0_{AB}-\delta^{AB}+2\delta^{A0}\delta^{B0},\nonumber\\
\end{eqnarray}
where $(ff)^0_{AB}=f_{0AC}f_{0CB}$ and so on, and all of them can be derived from the traceless nature of tensor operators and the symmetry of the constants $g_{ABC}$ and $f_{ABC}$.
Then, we obtain
\begin{eqnarray}
&&\Pi^{AB}_i=(gg)^0_{AB}+[\delta^{AB}-(gg)^0_{AB}]\tanh(\beta\phi^0_i)/(\beta\phi^0_i)\nonumber\\
&&\mbox{\hspace{1cm}}+g_{0AB}\tanh(\beta\phi^0_i),\;\;\;(A,B\ne 0).
\label{pola}
\end{eqnarray}

The saddle point $\langle \phi^0_{\mibs Q} \rangle$ satisfies the usual mean-field equation,
\begin{equation}
\phi\equiv\frac{\langle \phi^0_{\mibs Q} \rangle}{T_0}=\tanh(\beta\langle \phi^0_{\mibs Q} \rangle),
\end{equation}
where we have defined the upper transition temperature as $T_0=D(\mib Q)^{00}$.
In the lowest order we can replace $\phi_i^0$ by $\langle\phi^0_{\mibs Q}\rangle\exp(i{\mib Q}\cdot{\mib R}_i)$ in eq.~(\ref{pola}).
We replace every $\tanh(\beta\langle \phi^0_{\mibs Q} \rangle)$ by $\phi$, then the mode-coupling free energy $f_{\rm mc}$ is given by
\begin{eqnarray}
&&f_{\rm mc}=
\frac{1}{2}\sum_{\mibs q}\biggl[
\beta\phi^2\delta\phi^0_{\mibs q}\delta\phi^0_{-{\mibs q}}
-\beta\phi\sum_{AB}{}^\prime g_{0AB}\phi^A_{\mibs q}\phi^B_{-({\mibs q}+{\mibs Q})}
\nonumber\\&&\mbox{\hspace{5mm}}
+\sum_{AB}{}^\prime (\beta-T_0^{-1})\bigl(\delta_{AB}-(gg)^0_{AB}\bigr)\phi^A_{\mibs q}\phi^B_{-{\mibs q}}
\biggr],
\label{genfe}
\end{eqnarray}
in which $\delta\phi^0_{\mibs q}=\phi^0_{\mibs q}-\langle\phi^0_{\mibs Q}\rangle\delta_{\mibs q,Q}$.

In the case of $\langle \phi^0_{\mibs Q}\rangle = \langle \phi^{5gz}_{\mibs Q} \rangle$, the relevant symmetric constants are given by
\begin{eqnarray}
&&
(gg)^{15}_{AB}=\delta^{AB}\;\;\;(A=1,6,7,8,10,11),
\nonumber\\&&
g_{15,1,6}=1,\;\;g_{15,7,8}=g_{15,10,11}=-\sqrt{3}/2,
\nonumber\\&&
g_{15,7,11}=-g_{15,8,10}=1/2.
\end{eqnarray}
Putting them into eq. (\ref{genfe}), we obtain eq. (\ref{foxy}).

\section{Effective Hamiltonian in the Limit of $T_Q/T_N\gg 1$}
In this appendix, we discuss the effective Hamiltonian in the limit of $T_Q/T_N\gg 1$.
Let us introduce the $2\times 4$ matrix
\begin{equation}
U_i=P\exp(-i\pi\xi_i\tau^x_i/4),
\end{equation}
where the projection $P$ into $2\times2$-subspace of Kramers partners with the lower orbital energy is given by
\begin{equation}
P=\left(\begin{array}{cccc}
1 & 0 & 0 & 0 \\
0 & 0 & 0 & 1
\end{array}
\right).
\end{equation}
With this projection, the operators $X^A_i$ are transformed to
\begin{eqnarray}
&& X^{2u}_i\to \xi_i\sigma^z_i, \\
&& X^{4u1z}_i\to \sigma^z_i, \\
&& X^{4u2x}_i\to \frac{1}{2}(\sqrt{3}\sigma^x_i+\xi_i\sigma^y_i),\\
&& X^{4u2y}_i\to -\frac{1}{2}(\xi_i\sigma^x_i+\sqrt{3}\sigma^y_i),\\
&& X^{5ux}_i\to -\frac{1}{2}(\sigma^x_i-\sqrt{3}\xi_i\sigma^y_i),\\
&& X^{5uy}_i\to \frac{1}{2}(\sqrt{3}\xi_i\sigma^x_i-\sigma^y_i), \\
&& X^{5gz}_i\to \xi_i,
\end{eqnarray}
where ${\mib \sigma}$ is the two-dimensional Pauli matrix and the rest of operators are transformed to zero.

Then, the effective Hamiltonian is given in the following form:
\begin{equation}
H_{\rm eff}=-\frac{1}{2}\sum_{\mibs q}\sum_{\alpha\beta}
\left[S^{\alpha\beta}_{\mibs q}\sigma^\alpha_{\mibs q}\sigma^\beta_{-{\mibs q}}+T^{\alpha\beta}_{\mibs q}\sigma^\alpha_{\mibs q}\sigma^\beta_{-({\mibs q}+{\mibs Q})}\right],
\end{equation}
with
\begin{eqnarray}
&& S^{xx}_{\mibs q}=\frac{1}{2}\bigl(J_{5u}-J_{4u2}\bigr)j_0 \nonumber\\
&& \mbox{\hspace{2mm}} -\frac{1}{4}[(3K_{4u2}+K_{5u})j_x+(K_{4u2}+3K_{5u})j_y], \nonumber \\
&& S^{yy}_{\mibs q}=\frac{1}{2}\bigl(J_{5u}-J_{4u2}\bigr)j_0 \nonumber \\
&& \mbox{\hspace{2mm}} -\frac{1}{4}[(3K_{4u2}+K_{5u})j_y+(K_{4u2}+3K_{5u})j_x], \nonumber \\
&& S^{zz}_{\mibs q}=\bigl(J_{2u}-J_{4u1}\bigr)j_0-K_{4u1}j_z, \nonumber \\
&& S^{xy}_{\mibs q}=S^{yx}_{\mibs q}=-\bigl(K_{5u}-K_{4u2}\bigr)j_{xy}, \\
&& T^{xx}_{\mibs q}=T^{yy}_{\mibs q}=\frac{\sqrt{3}}{2}\bigl(K_{4u2}+K_{5u}\bigr)j_{xy}, \nonumber \\
&& T^{xy}_{\mibs q}=T^{yx}_{\mibs q}=\frac{\sqrt{3}}{4}\bigl(K_{5u}-K_{4u2}\bigr)\bigl(j_x+j_y\bigr).
\end{eqnarray}

Without $K_\Gamma$, the effective interactions are proportional to $J_{2u}-J_{4u1}$ or $J_{4u2}-J_{5u}$.
Then the lower transition temperature $T_N$ becomes zero in the case of the perfect competition, $J_{2u}=J_{4u1}$ and $J_{4u2}=J_{5u}$.
This is consistent with the discussions in \S3.2.

In the case of the perfect competition, the strength of the effective interactions by $K_\Gamma$ for ${\mib q}=(1/4,1/4,1/4)$ and for ${\mib q}=(1/4,1/4,0)$ are equal to each other.
Thus, the corresponding transition temperatures for both ${\mib q}$ become the same in the limit of $T_Q/T_N\gg 1$.
This is consistent with the results shown in Fig. \ref{fcnnn}.

\section{Legendre Transformation of Free Energy up to 4th Order}
In this appendix, we give the GL free energy in terms of the order parameter $\langle X\rangle$.
Adding the external field term $-\sum_iX^A_ih^A_i$ to the Hamiltonian (\ref{hamform}), the free energy in the partition function (\ref{part}) is given by
\begin{eqnarray}
&&{\cal F}=\frac{1}{2}{\sf D}^{-1}_{ab}(\phi^a-h^a)(\phi^b-h^b)-\frac{\beta}{2}\phi^a\phi^a-\frac{\beta^2}{3!}g_{abc}\phi^a\phi^b\phi^c \nonumber\\&&\mbox{\hspace{5mm}}
-\frac{\beta^3}{4!}(g_{abz}g_{zcd}-2\delta_{ab}\delta_{cd})\phi^a\phi^b\phi^c\phi^d+{\cal O}(\phi^5),
\end{eqnarray}
where $a=(A,i)$ etc., and the repeated indices should be summed.
The order parameter and the external field are related to each other by
\begin{equation}
h^a=\langle\phi^a\rangle-D_{ab}\langle X^b\rangle.
\end{equation}
If one uses the saddle-point approximation, the order parameter is expressed in terms of $\langle\phi\rangle$:
\begin{eqnarray}
&&\langle X^a\rangle=\beta\langle\phi^a\rangle+\frac{\beta^2}{2}g_{abc}\langle\phi^b\rangle\langle\phi^c\rangle
\nonumber\\&&\mbox{\hspace{5mm}}
+\frac{\beta^3}{3!}(g_{abz}g_{zcd}-2\delta_{ab}\delta_{cd})\langle\phi^b\rangle\langle\phi^c\rangle\langle\phi^d\rangle.
\end{eqnarray}
The converse relation is obtained as
\begin{eqnarray}
&&\beta\langle\phi^a\rangle=\langle X^a\rangle-\frac{1}{2}g_{abc}\langle X^b\rangle\langle X^c\rangle
\nonumber\\&&\mbox{\hspace{5mm}}
+\frac{1}{3}L_{abcd}\langle X^b\rangle\langle X^c\rangle\langle X^d\rangle,
\end{eqnarray}
where we have defined $L_{abcd}=g_{abz}g_{zcd}+\delta_{ab}\delta_{cd}$.
Then, the Legendre transform of the free energy, $G\equiv {\cal F}+\langle X^a\rangle h^a$, is given by
\begin{eqnarray}
G&&=\frac{1}{2}\sum_{\mibs q}\bigl(T\delta_{AB}-D^{AB}_{\mibs q}\bigr)\langle X^A_{\mibs q}\rangle \langle X^B_{-{\mibs q}}\rangle \nonumber\\
&&-\frac{T}{3!}\sum_{{\mibs q}{\mibs p}}g_{ABC}\langle X^A_{\mibs q}\rangle\langle X^B_{\mibs p}\rangle\langle X^C_{-({\mibs q}+{\mibs p})}\rangle\bigl.
\nonumber \\
&&+\frac{2T}{4!}\sum_{{\mibs q}{\mibs p}{\mibs k}}L_{ABCD}\langle X^A_{\mibs q}\rangle\langle X^B_{\mibs p}\rangle\langle X^C_{\mibs k}\rangle\langle X^D_{-({\mibs q}+{\mibs p}+{\mibs k})}\rangle\bigl.. \nonumber\\
\end{eqnarray}

%
%

%
%
\begin{table}
\caption{The multipole operators in the $\Gamma_8$ subspace.}
\begin{tabular}[t]{@{\hspace{\tabcolsep}\extracolsep{\fill}}cccc} \hline
$A$ & $\Gamma$ & symmetry & $X^A$ \\ \hline
 1 & $2u$  & $\sqrt{15}xyz$        & $\tau^y$ \\
 2 & $3g$  & $(3z^2-r^2)/2$        & $\tau^z$ \\
 3 &       & $\sqrt{3}(x^2-y^2)/2$ & $\tau^x$ \\
 4 & $4u1$ & $x$                   & $\sigma^x$ \\
 5 &       & $y$                   & $\sigma^y$ \\
 6 &       & $z$                   & $\sigma^z$ \\
 7 & $4u2$ & $x(5x^2-3r^2)/2$      & $\eta^+\sigma^x$ \\
 8 &       & $y(5y^2-3r^2)/2$      & $\eta^-\sigma^y$ \\
 9 &       & $z(5z^2-3r^2)/2$      & $\tau^z\sigma^z$ \\
10 & $5u$  & $\sqrt{15}x(y^2-z^2)/2$ & $\zeta^+\sigma^x$ \\
11 &       & $\sqrt{15}y(z^2-x^2)/2$ & $\zeta^-\sigma^y$ \\
12 &       & $\sqrt{15}z(x^2-y^2)/2$ & $\tau^x\sigma^z$ \\
13 & $5g$  & $\sqrt{3}yz$            & $\tau^y\sigma^x$ \\
14 &       & $\sqrt{3}zx$            & $\tau^y\sigma^y$ \\
15 &       & $\sqrt{3}xy$            & $\tau^y\sigma^z$ \\ \hline
\end{tabular}
\end{table}

\begin{table}
\caption{The antisymmetric constants $f_{ABC}$.}
\begin{tabular}[t]{@{\hspace{\tabcolsep}\extracolsep{\fill}}lclclc} \hline
$A,B,C$ & $f_{ABC}$ & $A,B,C$ & $f_{ABC}$ & $A,B,C$ & $f_{ABC}$ \\ \hline
1,2,3    & 1             & 1,7,10   & 1             & 1,8,11   & 1 \\
1,9,12   & 1             & 2,7,13   & $\sqrt{3}/2$  & 2,8,14   & $-\sqrt{3}/2$ \\
2,10,13  & $-1/2$        & 2,11,14  & $-1/2$        & 2,12,15  & 1 \\
3,7,13   & $1/2$         & 3,8,14   & $1/2$         & 3,9,15   & $-1$ \\
3,10,13  & $\sqrt{3}/2$  & 3,11,14  & $-\sqrt{3}/2$ & 4,5,6    & 1 \\
4,8,9    & $-1/2$        & 4,8,12   & $-\sqrt{3}/2$ & 4,9,11   & $-\sqrt{3}/2$ \\
4,11,12  & $-1/2$        & 4,14,15  & 1             & 5,7,9    & $1/2$ \\
5,7,12   & $-\sqrt{3}/2$ & 5,9,10   & $-\sqrt{3}/2$ & 5,10,12  & $1/2$ \\
5,13,15  & $-1$          & 6,7,8    & $-1/2$        & 6,7,11   & $-\sqrt{3}/2$ \\
6,8,10   & $-\sqrt{3}/2$ & 6,10,11  & $-1/2$        & 6,13,14  & 1 \\ \hline
\end{tabular}
\end{table}

\begin{table}
\caption{The symmetric constants $g_{ABC}$.}
\begin{tabular}[t]{@{\hspace{\tabcolsep}\extracolsep{\fill}}lclclc} \hline
$A,B,C$ & $g_{ABC}$ & $A,B,C$ & $g_{ABC}$ & $A,B,C$ & $g_{ABC}$ \\ \hline
1,4,13   & 1             & 1,5,14   & 1             & 1,6,15   & 1 \\
2,4,7    & $-1/2$        & 2,4,10   & $-\sqrt{3}/2$ & 2,5,8    & $-1/2$ \\
2,5,11   & $\sqrt{3}/2$  & 2,6,9    & 1             & 3,4,7    & $\sqrt{3}/2$ \\
3,4,10   & $-1/2$        & 3,5,8    & $-\sqrt{3}/2$ & 3,5,11   & $-1/2$ \\
3,6,12   & 1             & 7,8,15   & $-\sqrt{3}/2$ & 7,9,14   & $-\sqrt{3}/2$ \\
7,11,15  & $1/2$         & 7,12,14  & $-1/2$        & 8,9,13   & $-\sqrt{3}/2$ \\
8,10,15  & $-1/2$        & 8,12,13  & $1/2$         & 9,10,14  & $1/2$ \\
9,11,13  & $-1/2$        & 10,11,15 & $-\sqrt{3}/2$ & 10,12,14 & $-\sqrt{3}/2$ \\
11,12,13 & $-\sqrt{3}/2$ \\ \hline
\end{tabular}
\end{table}

%
%

\onecolumn

\begin{thebibliography}{99}
\bibitem{mfnote} The following empirical form for the solution is useful: $\xi=\tanh(1.85\sqrt{\beta T_Q-1})$.
\end{thebibliography}
\end{document}